\documentclass[aps,floats,amsmath,amssymb,showpacs,showkeys,superscriptaddress,preprint]{revtex4}
\usepackage{graphicx}
\usepackage{dcolumn}
\usepackage{hyperref}
\newcommand{\be}{\begin{equation}}
\newcommand{\ee}{\end{equation}}
\newcommand{\ba}{\begin{eqnarray}}
\newcommand{\ea}{\end{eqnarray}}
\begin{document}
\title{Phase diagram of softly repulsive systems: \\
The Gaussian and inverse-power-law potentials}
\author{Santi Prestipino}
\altaffiliation{Corresponding Author}
\email{santi.prestipino@unime.it}
\affiliation{Universit\`a degli Studi di Messina, Dipartimento di Fisica, Contrada Papardo, 98166 Messina, Italy}
\author{Franz Saija}
\email{saija@me.cnr.it}
\affiliation{Istituto per i Processi Chimico-Fisici del CNR, Sezione di Messina, Via La Farina 237, 98123 Messina,
Italy}
\author{Paolo V. Giaquinta}
\email{paolo.giaquinta@unime.it}
\affiliation{Universit\`a degli Studi di Messina, Dipartimento di Fisica, Contrada Papardo, 98166 Messina, Italy}
\date{\today}

\begin{abstract} ~~We redraw, using state-of-the-art methods for free-energy calculations, the phase diagrams of two
reference models for the liquid state: the Gaussian and inverse-power-law repulsive potentials. Notwithstanding
the different behavior of the two potentials for vanishing interparticle distances, their thermodynamic properties
are similar in a range of densities and temperatures, being ruled by the competition between the body-centered-cubic
(BCC) and face-centered-cubic (FCC) crystalline structures and the fluid phase. We confirm the existence of a reentrant
BCC phase in the phase diagram of the Gaussian-core model, just above the triple point . We also trace the BCC-FCC
coexistence line of the inverse-power-law model as a function of the power exponent $n$ and relate the common features
in the phase diagrams of such systems to the softness degree of the interaction.
\end{abstract}

\maketitle
%
%
\section{Introduction}
\setcounter{page}{1}
\renewcommand{\theequation}{1.\arabic{equation}
} A recurrent issue in statistical physics is the prediction of the phase behavior of a macroscopic system from the
knowledge of the force law which rules the interaction between the constituent particles. Over the last decades,
significant progress has been made in fulfilling this task, at least for simple model systems of classical point
particles interacting through a spherically-symmetric pair potential $v(r)$. In particular, continuous advancements in
both computer simulation methods and density-functional theory have revealed a few general features of the freezing
transition, a phase change that is prominently driven by the repulsive component of the interatomic potential. For
assigned thermodynamic conditions, the softness of the repulsion substantially determines the nature of the incoming
solid phase. In fact, a rigid potential favors a more compact phase, such as the face-centered-cubic (FCC) or the
hexagonal-close-packed (HCP) crystalline structure, as opposed to, say, the body-centered-cubic (BCC) one. This
is largely true even for more complex substances such as those that are generically classified as soft materials, like
colloidal suspensions for which the {\em effective} pair interactions can be accurately tailored through a fine
experimental control of suspended particle and solvent properties~\cite{Pusey,Likos1}.  A scholastic example is offered
by the entropy-driven repulsion between (the centers of mass of) self-avoiding polymer coils dispersed in a good
solvent~\cite{Likos1,Louis,Bolhuis}. In this specific case, the repulsion happens to be finite even if ``particles''
fully overlap and can be modelled with a Gaussian-shaped potential
\be v_{G}(r)=\epsilon\exp\left(-\frac{r^2}{\sigma^2}\right)\,.
\label{1-1}
\ee In Eq. (1.1) the length  $\sigma$ is of the order of the gyration radius of the polymer while the (positive) energy
$\epsilon$ is of the order of $k_{\rm B}T$, where $T$ is the temperature and $k_{\rm B}$ is the Boltzmann constant.
However, it is also useful to study this potential without assuming an explicit dependence of $\epsilon$ and $\sigma$
upon the temperature. The model is then referred to as the Gaussian-core model (GCM)~\cite{Stillinger1}. Other examples
of softly repulsive fluids are provided by dilute solutions of colloidal particles, sterically stabilized against
coagulation through a polymer coating. In this case, the strength and range of the effective repulsion between a
pair of particles are determined by the thickness of the layer of grafted polymer chains~\cite{Likos1} and can be
modelled with an inverse-power-law (IPL) potential
\be v_n(r)=\epsilon\left(\frac{\sigma}{r}\right)^n\,
\label{1-2}
\ee upon choosing a suitable power exponent $n$. In the past, this potential has been also used for a simplified
description of the thermal behavior of various simple metals under extreme thermodynamic conditions~\cite{Hoover1,Vocadlo}.

With the advent of numerically accurate techniques for the calculation of the Helmholtz free energy of a solid, computer
simulation has become a valuable tool for the determination of fluid-solid and solid-solid coexistence lines.
Most notably, the Frenkel-Ladd (FL) method can be used in conjunction with conventional thermodynamic integration
techniques to calculate the ``exact'' phase diagram of the system, the only source of error being ultimately associated
with the quality of the statistical sampling of the (finite) system~\cite{Frenkel1,Polson,Frenkel2}. Full-fledged
computer studies have been carried out for the repulsive Yukawa potential~\cite{Meijer,Hamaguchi}, typically used to
model solutions of charged colloids with counterions~\cite{Robbins,Sirota}, and for the logarithmic-plus-Yukawa potential
modelling star-polymer interactions~\cite{Witten,Likos2,Watzlawek}. In this paper, we present accurate results for the
phase diagrams of the Gaussian-core model and of the inverse-power-law potential, with the aim of providing a
comparable benchmark for such two potentials. A preliminary report on the present findings for the GCM has been already
given in~\cite{Prestipino}. The main results of this study are: i) the discovery of a reentrant BCC phase in the phase
diagram of the GCM; ii) an accurate determination of the BCC-FCC coexistence point of the IPL potential as a function of
the softness parameter $1/n$; iii) the formulation of an argument that accounts for the similarity of the
low-density/low-temperature behavior exhibited by the two models.
%
%
\section{Models and method}
\renewcommand{\theequation}{2.\arabic{equation}
}
\setcounter{equation}
{0}
\subsection{The Gaussian-core model} As is well known, the freezing transition of a many-particle system is ultimately
promoted by the Pauli repulsion between inner-shell electrons, causing the effective interatomic potential to diverge at
short distances. It is probably less known that the existence of a thermodynamically stable solid does not require a
singular repulsion for vanishing interatomic separations. In fact, even a finite square-barrier potential (penetrable
spheres) can support a stable solid phase at all temperatures~\cite{Likos3,Schmidt}. In this respect, the Gaussian
potential is a more realistic form of bounded repulsion. The GCM was originally introduced by
Stillinger~\cite{Stillinger1}. Two distinctive features of the phase diagram of the GCM, that are absent in other
bounded-interaction models, are the existence of a maximum melting temperature, $T_{\rm max}$, and,
correspondingly, of a reentrant melting transition into a denser fluid phase for $T<T_{\rm
max}$~\cite{Stillinger1,Stillinger2,Stillinger3,Lang}. In his original paper on the GCM,
Stillinger noted that, in the limit of vanishing temperature and density, the Gaussian particles behave as hard spheres
with a larger and larger diameter. In the same limit, the fluid freezes into a FCC structure whose number density $\rho$
vanishes with the temperature $T$ (from now on, temperature, pressure, and density will be given in reduced units, {\it
i.e.}, $T^*=k_{\rm B}T/\epsilon$, $P^*=P\sigma^3/\epsilon$, and $\rho^*=\rho\sigma^3$). The calculation of the total
energy of different cubic structures shows that the FCC structure is favored, at zero temperature, for reduced densities
lower than $\pi^{-3/2}\simeq 0.1796$. Beyond this threshold, a BCC solid takes over. However, upon compression,
any regular arrangement of particles is eventually doomed to collapse for any $T>0$ (reentrant
melting)~\cite{Stillinger2,Likos4}. An extended study of the GCM phase diagram was carried out by Lang and coworkers who
employed an integral-equation theory for the fluid phase and a harmonic approximation for the crystalline
phases~\cite{Lang}. In this combined approximation, these authors estimated a maximum melting temperature $T_{\rm
max}^*=0.0102$ at which the coexisting fluid and BCC phases have a reduced density $\rho_{\rm max}^{*}=0.2292$. They
also found that the FCC-BCC coexistence lines run {\it linearly} from the points $\rho_{\rm FCC}^{*}=0.17941$ and
$\rho_{\rm BCC}^{*}=0.17977$ at $T^{*}=0$ to the points $\rho_{\rm FCC}^{*}=0.16631$ and $\rho_{\rm BCC}^{*}=0.16667$ at
the triple-point temperature $T_{\rm tr}^*=0.008753$.

\subsection{The inverse-power-law fluid} The IPL potential has been the subject of many
studies~\cite{Hoover2,Laird,Agrawal}. This potential provides a continuous path from hard spheres ($n\rightarrow\infty$)
to the one-component plasma ($n=1$). Actually, a neutralizing background is needed for $n\leq 3$ to balance the
non-integrable long-ranged repulsion~\cite{Brush,Hansen}. The thermodynamic properties of the model can
be expressed in terms of just one quantity, $\gamma=\rho^*T^{*\,-3/n}$, as a result of the scaling properties of the
excess Helmholtz free energy. Hence, for any given $n$, it is sufficient to trace the thermal behavior of the IPL fluid
as a function of $\rho^*$ along the isotherm $T^*=1$.  For large values of $n$, the BCC phase is found to be unstable
with respect to shear modes. As a result, the fluid freezes into a FCC solid. As $n$ decreases, the potential becomes
increasingly softer and longer ranged until, for $n\approx 8$, the BCC phase becomes mechanically
stable. For smaller values of $n$ and for densities close to freezing, the entropy of the BCC solid turns out to be
greater than that of the FCC phase since the former structure can actually sustain a larger number of soft
shear modes. However, the FCC phase is favored at high densities and low temperatures because of its smaller energy.
Hence, upon reducing $\gamma$, a FCC-BCC transition may actually anticipate the melting of the solid for intermediate
values of $n$. Agrawal and Kofke located the triple point in the range $0.11 \lesssim 1/n \lesssim 0.17$, further
claiming that a more realistic estimate of the lower bound, based on the mechanic stability range of the BCC phase, was
$0.1305$~\cite{Agrawal}. However, they did not provide a direct and systematic estimate of the location of the FCC-BCC transition as a function of
the power exponent $n$.

\subsection{Details of the Monte Carlo simulation} We performed Monte Carlo (MC) simulations of the GCM and of the IPL
model in the canonical ensemble, implementing the standard Metropolis algorithm with periodic boundary conditions and
with the nearest-image convention. The numbers of particles $N$ was chosen so as to fit a cubic box of volume $V$ with an
integer number of cells: $N=4m^3$ for the FCC solid and $N=2m^3$ for the BCC solid, $m$ being the number of
cells along any spatial direction. The nearest-neighbor distance $a$ is $(\sqrt{2}/2)(L/m)$ for a FCC lattice and
$(\sqrt{3}/2)(L/m)$ for a BCC lattice, where $L=(N/\rho)^{1/3}$. Our typical samples consisted of 1372 particles in the
fluid phase and in the solid FCC phase, and of 1458 particles in the BCC phase. We emphasize that, unless one resorts to
some extrapolation to infinite size (a computationally demanding task), one should compare the statistical
properties of samples with similar sizes, in particular if one wants to investigate the relative thermodynamic
stability of different phases at given temperature and pressure. Occasionally, we considered smaller as well as
larger sizes to check whether our results could be possibly affected by a significant size dependence (and they were
actually not). We also carried out a number of simulations of a HCP solid hosting $10\times 12\times 12=1440$
particles.

In order to locate the freezing point at a given temperature, we generated distinct chains of simulations, starting from
the very dilute fluid on one side and from the dense solid on the other side. As a rule, the last MC configuration
produced  for a given value of $\rho$ was used, after a suitable rescaling of particle coordinates, as the starting point
for the next run at a slightly higher or lower density. We monitored the difference in energy and pressure between the
various phases until we observed a sudden change of both quantities, so as to avoid averaging over heterogeneous
thermodynamic states. Usually, the solid can be overheated for a few more steps beyond the melting point. The
undercooling of the fluid is notoriously much easier. For any $\rho$ and $T$, the equilibration of each sample typically
took a few thousand MC sweeps, a sweep consisting of one attempt to change sequentially the position of all the
particles. The maximum random displacement of a particle in a trial MC move was adjusted once a sweep during the run so
as to keep the acceptance ratio as close to 50\% as possible, only small excursions around this value being allowed.
Thermodynamic averages were computed over trajectories $3\times 10^4$ sweeps long. The pressure was computed through the
virial formula \be P=\rho k_{\rm B}T+\frac{1}{3V}{\left<{-\sum_{i<j}r_{ij}v'(r_{ij})}\right>}\,, \label{2-1} \ee where
$r_{ij}$ is the distance between particles $i$ and $j$ and $\left<\cdots\right>$ denotes the canonical average. For the
IPL fluid, the excess pressure is proportional to the excess energy $u_{\rm ex}$ since $-rv'_n(r)=nv_n(r)$. To avoid
counting pair interactions twice, the potential was truncated at a cutoff distance $r_c$, which was taken to
be just slightly smaller than $L/2$. Correspondingly, long-range corrections to the energy and pressure were calculated
upon assuming the radial distribution function (RDF) $g(r)$ to be unity for $r>r_{c}$. The histogram of $g(r)$ was
constructed with a spatial resolution $\Delta r=\sigma/50$ and was updated every 10 MC sweeps. The RDF was never
found to differ significantly from unity -- in any phase -- at the maximum distance $L/2$, at least for the largest sample sizes.

The numerical estimate of statistical errors is a critical issue whenever different candidate solid structures so closely
compete for thermodynamic stability. To this aim, we divided the MC trajectory into ten blocks and estimated the
error bars to be twice as large as the standard deviation of the block averages, under the implicit assumption that the
decorrelation time of any relevant variable was less than the size of a block. Tipically, the relative errors affecting
the energy and pressure of the fluid were found to be a few hundredths percent at the most (for a solid, they were even
smaller). We computed the difference between the excess free energies of any two equilibrium states (say, 1 and 2)
belonging to the same phase with the standard thermodynamic integration technique, {\it i.e.}, via the combined use of
the formulae
\be
\frac{f_{\rm ex}(\rho,T_2)}{T_2}=\frac{f_{\rm ex}(\rho,T_1)}{T_1}-
\int_{T_1}^{T_2}{\rm d}T\,\frac{u_{\rm ex}(\rho,T)}{T^2}
\label{2-2}
\ee and
\be
\beta f_{\rm ex}(\rho_2,T)=\beta f_{\rm ex}(\rho_1,T)+
\int_{\rho_1}^{\rho_2}{\rm d}\rho\,\frac{1}{\rho}
\left[\frac{\beta P(\rho,T)}{\rho}-1\right]\,.
\label{2-3}
\ee In the above equations, $f_{\rm ex}$ is the excess Helmholtz free energy per particle and $\beta=(k_{\rm B}T)^{-1}$.
The integrals in Eqs. (2.2) and (2.3) were performed numerically with Simpson's rule over a cubic-spline approximant of
the simulation data. The excess values of the chemical potential and of the entropy per particle follow from
\be
\beta\mu_{\rm ex}=\beta f_{\rm ex}+\frac{\beta P}{\rho}-1
\,\,\,\,\,\,{\rm and}\,\,\,\,\,\,
\frac{s_{\rm ex}}{k_{\rm B}}=\beta(u_{\rm ex}-f_{\rm ex})\,.
\label{2-4}
\ee Equations (2.2) and (2.3) are useless if an independent estimate of the free energy is not available for
some selected reference state of each phase. The free energy of a solid can be obtained with the FL method, that is
reviewed in the next paragraph. As for the fluid, a reference state can be any equilibrium state characterized by a very
small density (a nearly ideal gas), since then the excess chemical potential $\mu_{\rm ex}$ can be accurately
estimated through Widom's particle-insertion method~\cite{Widom,Frenkel2}
\be
\mu_{\rm ex}=-k_{\rm B}T\ln\left<\exp\left(-\beta E_{\rm ins}\right)\right>\,,
\label{2-5}
\ee where $E_{\rm ins}$ includes all the interactions between a randomly inserted test particle and the particles of
the fluid. Typically, the average in Eq. (2.5) was computed over equilibrium runs $6\times 10^4$ sweeps long, an
insertion being attempted at the completion of each sweep.

The chemical potential is the quantity that decides which phase is stable for given values of $T$ and $P$. Hence, it is
necessary to inspect the error affecting this quantity with care. Let us consider, for instance, the chemical potential
of a solid whose density is $\rho_2$ in a state that is connected, along an isothermal path, to another (reference) state
with density $\rho_1$. Let $\delta f_1$ be the numerical precision of the FL calculation of the free energy of the
reference solid. For each $NVT$ simulation carried out along the isothermal path, one can estimate the error affecting
the compressibility factor $\beta P/\rho$, whence its mean error  $\delta z$ along the path. Hence, according to Eqs.
(2.3) and (2.4), the error affecting $\mu$ is approximately given by $\delta f_1+k_{\rm B}T\delta z(1+\ln(\rho_2/\rho_1))$.

\subsection{The Frenkel-Ladd calculation of the solid free energies} The method proposed by Frenkel and Ladd
to calculate the free energy of a solid relies on a different kind of thermodynamic integration~\cite{Frenkel2}. The idea
is to connect continuously the state of a solid (say, system $1$ with potential energy $V_1$), to the state of another
solid (say, system $0$ with potential energy $V_{0}$) whose free energy is known. Upon assuming the reference solid to be
an Einstein solid with spring constant $c$, a hybrid system is sampled with potential $V_\lambda=V_0+\lambda(V_1-V_0)$
($0\leq\lambda\leq 1$), for given $\rho$ and $T$. The Helmholtz free energy of system $1$ is then obtained through
Kirkwood's formula:
\be F_1-F_0=\int_0^1{\rm d}\lambda\left<V_1-V_0\right>_\lambda\,.
\label{2-6}
\ee The free energy of the Einstein solid and the mean square separation of an Einstein particle from its reference
lattice site are:
\be
\beta F_0=-\frac{3N}{2}\ln\left(\frac{2\pi}{\Lambda^2\beta c}\right)\,,
\label{2-7}
\ee
\be
\Delta R^2_0=\frac{3}{\beta c}\,,
\label{2-8}
\ee where $\Lambda$ is the thermal wavelength. A nontrivial problem with this choice of $V_0$, already pointed out in
the original paper \cite{Frenkel1}, is the following: The Einstein particles can only perform limited excursions
about their reference lattice positions, which implies a finite value of $\Delta R^2$; however, there is no means to
constrain particles interacting with the potential $V_1$ to move within the neighborhood of their initial positions, even
in the solid phase. In other words, at variance with $V_0$, $V_1$ is a translationally-invariant potential, which implies
that the integrand in Eq. (2.6) diverges for $\lambda=1$. This problem can be overcome by performing MC simulations of
$V_\lambda$ under the constraint of a fixed center of mass. Only afterwards, the proper corrections are taken into
account~\cite{Polson}. We just quote the final result:
\ba
\beta f_{\rm ex}(N) &=& \frac{\beta V_1^{(0)}}{N}-
\ln\left(\rho(\beta c)^{-3/2}\right)-\frac{3}{2}\ln(2\pi)+1-
\frac{2\ln N}{N}+\frac{\ln(2\pi)}{N}+
\frac{\ln\left(\rho(\beta c)^{-3/2}\right)}{N}
\nonumber \\
&+& \frac{\beta}{N}\int_0^1{\rm d}\lambda
\left<V_1-V_1^{(0)}-V_0\right>_\lambda^{\rm CM}\,, \label{2-9} \ea where the superscript CM denotes a constrained
average and $V_1^{(0)}$ is the total potential energy $V_1$ for particles located in their respective lattice positions.
A linear scaling relation was conjuctered, on rather general grounds, for the asymptotic behavior of the free energy as a
function of the sample size: $\beta f_{\rm ex}(N)+\ln N/N=\beta f_{\rm ex}(\infty)+{\cal O}(N^{-1})$~\cite{Polson}.

We finally add some further considerations about the numerical implementation of the FL method. In principle, one might
choose the value of $c$ in an arbitrary way. However, if the quantity $\Delta R^2$ for the target solid is close to $3/(\beta
c)$, the variations of the integrand in Eq. (2.9) and, correspondingly, the error made in the numerical
quadrature get very small. In practice, $\lambda$ was taken to increase from 0 to 0.9 with steps of 0.05, and with a
smaller step $(0.01)$ in the range from 0.9 to 1. For each value of $\lambda$, as many as $2\times 10^4$ MC sweeps were
produced at equilibrium. The MC moves were such that the position of the center of mass of the solid was held fixed;
this way, particles can undergo only limited excursions along the run. For this reason, if a particle happens to move
out of the simulation box, there is no need to put it back.
%
%
\section{Results}
\renewcommand{\theequation}{3.\arabic{equation}
}
\setcounter{equation}
{0}
\subsection{Gaussian-core model}
Figure \,1 presents the phase diagram of the GCM, calculated with the numerical techniques discussed above, in the
temperature-pressure plane. The general features of this phase diagram have already been illustrated in a number of
papers~\cite{Stillinger3,Lang,Stillinger4,Giaquinta}. A novel feature that has been just reported is the emergence, at
low densities and in a narrow range of temperatures ($0.0031 \lesssim T^* \lesssim 0.0037$), of a BCC phase
intermediate between the fluid and the FCC phase~\cite{Prestipino}. Correspondingly, the thermodynamic coordinates of the
triple point are: $T_{\rm tr}^*=0.0031$ and $P_{\rm tr}^*=0.0070$. Calculations carried out with different number of particles
confirm that the reentrant BCC phase at low densities is not a finite-size effect (see the inset in Fig.\,1). Tracing the coexistence lines in a
phase diagram requires the calculation of the chemical potential, $\mu(T,P)$, for all candidate phases. In the case of
the GCM, the reentrance of the fluid phase at high densities for $T<T_{\rm max}$, calls for an accurate calculation of
this quantity in the dense-fluid regime. However, in such thermodynamic conditions Widom's method is no longer reliable.
We thus resorted to a thermodynamic integration along a composite path in the fluid-phase region: We first moved from low to
high densities ($\rho^*=0.74$) at fixed temperature ($T^*=0.015$), and followed therefrom an isochoric path down to low
temperatures ($T^*=0.002$). Differences between the chemical potentials of fluid and solid phases were plotted as a
function of the pressure in Figs.\,2 and \,3. The sequence of phases showing up with increasing pressures at $T^*=0.002$
is fluid, FCC, BCC, fluid again. It is also apparent from Fig.\,2 that the chemical-potential gap between the FCC and BCC
phases depends on the pressure in a distinctly nonmonotonic way: At low pressures, the BCC phase already competes with
the FCC phase and is about to become stable. The changeover from the FCC to the BCC phase occurs at higher temperatures.
In fact, the shallow valley in the quantity $\mu_{\rm FCC}-\mu_{\rm BCC}$ gradually slides upwards with the temperature
until, for $T^*\gtrsim 0.0030$, the BCC phase slips in between the fluid and the FCC phases. As also seen
from Fig.\,3, the HCP phase never comes into play.

The excess Helmholtz free energies calculated for some solid phases of the GCM are given in Table 1, together with the
values of the reduced elastic constant $c^*=c\sigma^2/\epsilon$. Though one single FL calculation would be sufficient to
estimate the free energy of any given phase in any other thermodynamic state, we found it more practical to repeat the
calculation rather than generating a dense mesh of thermodynamic paths. As illustrated in Fig.\,4, the free energy scales
asymptotically with $N$ as conjectured by Polson and coworkers~\cite{Polson}.

The thermodynamic properties of the GCM at fluid-solid and solid-solid coexistence are given in Table 2. Upon
increasing the pressure for temperatures just above the triple point, the entropy per particle
drops by about $0.8\,k_{\rm B}$ and $0.1\,k_{\rm B}$ across the fluid-to-BCC and BCC-to-FCC phase transitions,
respectively. However, the next jump at the FCC-to-BCC transition is positive in that the entropy of the coexisting BCC
phase, notwithstanding its slightly higher density (see Table 2), again exceeds that of the FCC solid by about the same
amount $(0.1\,k_{\rm B})$. As then implied by the Clausius-Clapeyron equation, the slope of the higher-pressure branch
of the FCC-BCC coexistence line is negative in the $(T,P)$ plane as is also the slope of the upper branch of the
BCC-fluid coexistence line. In fact, at the reentrant melting point, the coexisting fluid has both a larger density and
a larger entropy than the BCC phase, the entropy jump being again of the order of  $0.8\,k_{\rm B}$ per particle.

We actually found that for low densities ($\rho^{*} \lesssim 0.2$) the entropy of the BCC solid
systematically overcomes that of the FCC solid, even below the triple-point temperature. The gap decreases slowly with
increasing temperatures and is likely due to the larger number of soft shear modes hosted by the BCC structure.
Actually, even at low temperatures, BCC-ordered grains are expected to form in the freezing fluid, a circumstance which
may slow down the crystallization process. On the energy side, the scales tip instead in favor of the FCC solid. In
fact, at low densities ($\rho^* \lesssim 0.17$), the FCC phase has a lower energy than the BCC phase for $T^{*} \lesssim
0.008$. Not surprisingly, this temperature is very close to the reduced triple-point temperature predicted by Lang and
coworkers~\cite{Lang}. These authors actually used the Gibbs-Bogoliubov inequality to optimize a harmonic model of both
solid phases. Considering the fact that the FCC-BCC transition occurs at rather low densities, the above approximation
does likely underestimate the (relative) entropies of the competing crystalline phases. As a result, the FCC phase turns
out to be stable in a much more extended region than that predicted by free-energy calculations carried out with the FL
method (see also Fig.\,2 of~\cite{Prestipino}).

As shown in Fig.\,1, we predict that the BCC phase is stable for temperatures lower than $T_{\rm max}^*\simeq 0.00874$.
The reduced density and pressure of the solid coexisting with the fluid at this temperature are $\rho_{\rm max}^*\simeq
0.239$ and $P_{\rm max}^*\simeq 0.128$, respectively. At this bending point the BCC solid melts, upon heating, with no
volume change. However, the transition is still first-order since the entropy per particle of the fluid is $0.79\,k_{\rm
B}$ larger than in the solid. As for the FCC phase, our current estimate for the uppermost FCC-BCC coexistence
temperature is $0.0038$. The corresponding estimates for the reduced density and pressure attained by the system in this
extremal state of the FCC stability basin are $0.123$ and $0.0174$, respectively. Comments analogous to those made above
on the nature of the BCC-fluid phase transition at the maximum temperature of the coexistence line can be straightly
extended to the FCC-BCC transition as well.

Figure\,5 shows the RDFs of the fluid and solid phases for $T^*=0.004$, at both ordinary and reentrant melting densities.
We note that neighboring particles lie, on average, at relative distances larger than $\sigma/\sqrt{2}$, which is where
the Gaussian potential inflects.

\subsection{Inverse-power-law model}
We list the excess Helmholtz free energies for some crystalline states of the IPL model in Table \,3. Figure \,6 shows
the chemical potential plotted as a function of the pressure for an intermediate value of the softness parameter,
$1/n=0.16$. Upon increasing $\gamma$, the IPL fluid freezes into a BCC structure. A further compression triggers a
solid-solid transition into a FCC structure. Correspondingly, the equation of state (Fig.\,7) shows two tie lines
joining the three distinct branches at the fluid-BCC and BCC-FCC transition points. We found that the HCP solid is
systematically less stable than the FCC solid, although the chemical-potential gap between the two phases is rather
small at all pressures. As $n$ grows, the range of BCC stability gradually reduces until, for $1/n\lesssim 0.14$, the
FCC structure only is left over. The maximum value of the power exponent for which the BCC solid is thermodynamically
stable is close to 7 $(0.14 \lesssim 1/n < 0.15)$.The phase diagram of the IPL fluid is shown in Fig.\,8 while the
phase-transition densities are listed in Table \,4. In the triple-point region, the density jumps are approximately
$0.030$ across the fluid-BCC transition and $0.004$ across the BCC-FCC transition; correspondingly, the entropy decreases
at each transition by about $0.7\,k_{\rm B}$ and $0.1\,k_{\rm B}$ per particle, respectively. All along the freezing
line, the entropy of the BCC solid overcomes that of the FCC solid by approximately $0.1\,k_{\rm B}$ per particle. We
verified that the estimate of the power exponent $n_{\rm tr}$ for which the IPL model exhibits triple-phase coexistence is not affected by
appreciable finite-size effects because of the large samples used, the only significant source of error being the limited
statistics of the simulation. The maximum error expected on $\beta\Delta\mu$, near to freezing, is approximately equal to
$1.5\times 10^{-2}$, as for the GCM~\cite{Prestipino}. Such an error would, in turn, affect the estimates of the reduced
phase-transition densities by an amount not greater than $3\times 10^{-2}$. We conclude that the error on $n_{\rm tr}$
should not be larger than $0.5$.

Figure \,8 presents the phase diagram of the IPL model, spanned by the softness parameter $1/n$ and by the scaled
quantity $\gamma$. We also plotted fluid-solid coexistence data obtained by other authors using different
approaches~\cite{Hoover2,Agrawal,Laird,Wang}. The data reported in~\cite{Hoover2} refer to a single-occupancy solid
in which each particle is confined within a spherical cell whose diameter is set equal to the FCC nearest-neighbor
lattice spacing. Dubin and Dewitt estimated the fluid-solid and solid-solid coexistence of the IPL model by applying the
empirical Lindemann criterion, including some anharmonic contributions to the internal energy of the solids~\cite{Dubin}.
Their fluid-FCC coexistence line almost coincides with the present fluid-solid coexistence line; instead, the location of
the BCC threshold does not agree with the present data.
%
%
\section{Discussion}
\renewcommand{\theequation}{4.\arabic{equation}
} \setcounter{equation}
{0} The low-density/low-temperature topology of the phase diagram of the GCM, with a triple point separating a region
where the fluid freezes into a FCC structure from another region where such two phases are bridged by an intermediate
BCC phase, is common to other model systems with softly repulsive interactions. Another example, besides the GCM and the
IPL model, is the Yukawa potential, $v_{\rm Y}(r)=B\exp(-r/\ell)/r$ (with $B>0$), which has been used to model the
thermal behavior of dilute solutions of charged colloids with counterions~\cite{Meijer,Hamaguchi,Robbins}. The phase
diagram of this latter model can be conveniently spanned by two scaled quantities, {\it i.e.}, $\rho\ell^3$ and
$B\rho^{1/3}/(k_{\rm B}T)$~\cite{Hamaguchi}. In an attempt to relate the phase behaviors of the Gaussian and Yukawa
potentials to that of the IPL model, we required in~\cite{Prestipino} the logarithmic derivatives of the three potentials to
match, at least for separations close to the average nearest-neighbor distance, $\overline{r}=\rho^{-1/3}$. This
condition yields two ``optimal'' values for the IPL power exponent: $n_{\rm G}=2(\rho\sigma^3)^{-2/3}$ and $n_{\rm
Y}=1+(\rho\ell^3)^{-1/3}$, respectively. Increasing the density of the Gaussian and Yukawa models along the freezing
line is tantamount, through the above relations, to an increase of the softness parameter $1/n$ of the ``equivalent" IPL
model. In both cases the system is driven towards the BCC phase. The values of $n_{\rm G}$ and $n_{\rm Y}$ calculated at
the triple point of the Gaussian and Yukawa potentials are $9.6$ and $12.0$, respectively, not too far from the IPL value
$(\approx 7)$. This rough argument shows that it is possible to gauge the softness of such potentials in such a way that
their phase behavior is referred to a common frame where the BCC phase is ultimately promoted by a sufficiently
soft interaction.

A similar interplay between BCC and FCC solid phases is also present in the phase diagram of the modified Buckingham
potential that is used to model the behavior of rare gases at high temperature and pressure~\cite{Saija}: under extreme
thermodynamic conditions close to freezing, notwithstanding the presence of an attractive tail, particles feel almost
exclusively the repulsive shoulder of the potential, whose softness increases with the temperature.
%
%
\section{Conclusions} In this paper we revisited the phase diagram of two classical reference models for the liquid
state, both governed by softly repulsive interactions with Gaussian and inverse-power-law shapes, respectively. The aim
of this effort was to point out some aspects of the phase diagrams which had been previously overlooked or simply
ignored, in a common theoretical framework for repulsive pair potentials that decrease polynomially or faster with
the interparticle distance. To this end, we carried out extensive Monte Carlo simulations and calculated the
thermodynamic properties of both fluid and solid phases through the combined use of thermodynamic integration techniques
and free-energy calculations. We further discussed the similarities and semi-quantitative correspondences in the phase
diagrams of such models in relation to the softness of the potential and with specific reference to the interplay of
fluid, BCC and FCC crystalline phases.
%
%

\newpage
%
%
\begin{center}
\large TABLE CAPTIONS
\normalsize
\end{center}
\begin{description}
\item[{\bf Table 1 :}] Excess Helmholtz free energy per particle $f_{\rm ex}$, in units of $k_{\rm B}T$, for some
crystalline states of the GCM: FCC ($N=1372$), HCP ($N=1440$), and BCC ($N=1458$). For $T^*=0.003$ and $0.006$, the
tabulated values refer to systems with $864$ and $1024$ particles, respectively. The value of the reduced elastic
constant $c^*=c\sigma^2/\epsilon$, playing a role in the Frenkel-Ladd calculation, is also displayed in square brackets:
for such values of $c$, the mean square displacement of the Einstein solid approximately matches the mean square
deviation of a GCM particle from its reference lattice site.

\item[{\bf Table 2 :}] Thermodynamic properties of the GCM at fluid-solid and solid-solid coexistence. With the only
exception of the states identified by $T^*=0.003$ and $0.006$, the simulation data refer to samples with $N=1372$ (fluid
and FCC) and $N=1458$ particles (BCC). For $T^*=0.003$ and $0.006$, the tabulated values are relative to samples with
$864$ and $1024$ particles, respectively. We show, from left to right, the values relative to the following first-order
phase transitions: freezing of the fluid into either a FCC or BCC solid; BCC-to-FCC transition; FCC-to-BCC transition,
and high-density melting of the BCC solid. We indicated with $\rho_{\rm f}$ and $\rho_{\rm m}$ the freezing and melting
densities, respectively. All data were arbitrarily rounded off; note, however, that the error originating from the
limited precision of both the Monte Carlo data {\em and} the free-energy values may be actually larger than that
apparently implied by the chosen truncation.

\item[{\bf Table 3 :}] Excess Helmholtz free energy per particle $f_{\rm ex}$, in units of $k_{\rm B}T$, for some
crystalline states of the IPL model: FCC ($N=1372$), HCP ($N=1440$), and BCC ($N=1458$). The value of the reduced
elastic constant $c^*=c\sigma^2/\epsilon$ is also displayed in square brackets (see the caption of Table 1).

\item[{\bf Table 4 :}] Thermodynamic properties of the IPL model at fluid-solid and solid-solid coexistence for some
values of the softness parameter $1/n$. The tabulated data refer to samples with $N=1372$ (fluid, FCC) and $N=1458$
particles (BCC). All data were arbitrarily rounded off at the third decimal digit (see the caption of Table 2).

\end{description}

\newpage
%
%
\begin{center}
\large FIGURE CAPTIONS
\normalsize
\end{center}
\begin{description}

\item[{\bf Fig.\,1 :}] Phase diagram of the GCM in the $(T,P)$ plane: The markers (open circles) identify the states
were we carried out the numerical calculations with $N=1372$ particles for the fluid and FCC phases and with $N=1458$ for
the BCC phase. The triple-point region is magnified in the inset: The solid circles mark the phase thresholds in larger
samples with $2048$ particles for the fluid and FCC phases and with $2000$ particles for the BCC phase, at $T^*=0.0037$.
The FCC-BCC coexistence pressure at $T^{*}=0$ (open square) was taken from~\cite{Stillinger1}. The lines drawn through
the data points are a guide for the eye.

\item[{\bf Fig.\,2 :}] Chemical-potential differences between pairs of GCM phases plotted as a function of the pressure
for $T^*=0.002$: $\beta\Delta\mu_{\rm fluid,\,FCC}$ (dashed line); $\beta\Delta\mu_{\rm fluid,\,BCC}$ (dotted line);
$\beta\Delta\mu_{\rm FCC,\,BCC}$ (continuous line). Upon increasing either $P$ or $\rho$, the GCM exploits a
fluid-FCC-BCC-fluid sequence of phases. A zoom in the low-pressure region reveals the non-monotonic behavior of
$\beta\Delta\mu_{\rm FCC,\,BCC}$, a feature that is ultimately responsible for the reentrance of the BCC phase at higher
temperatures (see also Fig. 3). The lines are spline interpolants of the data.

\item[{\bf Fig.\,3 :}] Chemical potential differences between pairs of GCM phases plotted as a function of the pressure
for $T^*=0.0037$: $\beta\Delta\mu_{\rm fluid,\,FCC}$ (dashed line), $\beta\Delta\mu_{\rm fluid,\,BCC}$ (dotted line),
$\beta\Delta\mu_{\rm FCC,\,HCP}$ (dash-dotted line), and $\beta\Delta\mu_{\rm FCC,\,BCC}$ (continuous line). The lines
for $\beta\Delta\mu_{\rm fluid,\,FCC}$ and $\beta\Delta\mu_{\rm fluid,\,BCC}$ are hardly resolved on the scale of the
figure. Upon increasing the pressure, the GCM exploits a fluid-BCC-FCC-BCC-fluid sequence of phases. The lines are
spline interpolants of the data.

\item[{\bf Fig.\,4 :}] Frenkel-Ladd calculation of the free energy of the GCM solid. Top: The integrand of
Eq. (2.9) plotted as a function of $\lambda$ for a BCC solid with $1458$ particles ($\rho^*=0.3$, $T^*=0.004$);
the continuous line is a spline interpolant of the data, the error bars being much smaller than the size of the symbols.
Bottom: The quantity $\beta f_{\rm ex}(N)+\ln N/N$ plotted as a function of $N^{-1}$ for a BCC solid in the same
thermodynamic state, with $N=432,686,1024,1458$. We checked the asymptotically linear scaling of the above quantity also
for other states among those listed in Table 1.

\item[{\bf Fig.\,5 :}] Radial distribution functions of fluid and solid phases of the GCM for $T^*=0.004$, $\rho^*=0.11$
(top) and $\rho^*=0.52$ (bottom): fluid (continuous line), BCC (dotted line), FCC (dashed line).

\item[{\bf Fig.\,6 :}] Chemical-potential differences between pairs of phases of the IPL fluid for $1/n=0.16$, plotted
as a function of the pressure for $T^*=1$: $\beta\Delta\mu_{\rm fluid,\,FCC}$ (dashed line); $\beta\Delta\mu_{\rm
fluid,\,BCC}$ (dotted line); $\beta\Delta\mu_{\rm FCC,\,HCP}$ (dash-dotted line); $\beta\Delta\mu_{\rm FCC,\,BCC}$
(continuous line). For increasing pressures, the IPL fluid exploits a fluid-BCC-FCC sequence of phases. The lines are
spline interpolants of the data.

\item[{\bf Fig.\,7 :}] Equation of state of the IPL model for $T^*=1$ and $1/n=0.16$. The fluid branch ($\bigcirc$,
continuous line) and the FCC branch ($\Box$, dashed line) were plotted for $N=1372$ particles, while the BCC branch
($\triangle$, dotted line) was obtained with a sample of 1458 particles. The coexistence densities, signalled by two
pairs of thin vertical lines, are located where the tie lines cross different branches of the equation of state.

\item[{\bf Fig.\,8 :}] Phase diagram of the IPL potential: fluid ($N=1372$, open circles); FCC ($N=1372$, open squares);
BCC ($N=1458$, open triangles). The continuous lines drawn through
the data are a guide for the eye. The dashed lines are the phase
thresholds reported in~\cite{Wang}. In the inset, we compare the
present data (continuous lines) with the phase-transition loci
reported in~\cite{Hoover2} ($n=6$: fluid, solid circle; FCC, solid
square); in~\cite{Laird} ($n=6$: fluid, star; BCC, tripod; FCC,
cross); and in~\cite{Agrawal} (fluid, cross; solid, tripod; note
that this latter datum refers to FCC for $1/n<0.16$ and to BCC
otherwise).
\end{description}

\newpage
%
%
\begin{center}
\large TABLE 1 \normalsize

\vspace{5mm}

\begin{tabular*}{\columnwidth}[c]{@{\extracolsep{\fill}}|r|r||r|r|r|}
\hline $\rho^*$ & $T^*$ & FCC\,\,\,($N=1372$) & HCP\,\,\,($N=1440$) & BCC\,\,\,($N= 1458$) \\
\hline\hline $\,0.30\,$ & $0.0020$ & 195.703(2)\,\,\,\,$[0.30]$\,\, & & 195.312(1)\,\,\,\,$[0.45]$\,\, \\
\hline $\,0.24\,$ & $0.0030$ & 86.251(2)\,\,\,\,$[0.35]$\,\, & & 86.057(1)\,\,\,\,$[0.39]$\,\, \\
\hline $\,0.24\,$ & $0.0033$ & 78.994(1)\,\,\,\,$[0.34]$\,\, & & 78.814(1)\,\,\,\,$[0.38]$\,\, \\
\hline $\,0.24\,$ & $0.0035$ & 74.835(1)\,\,\,\,$[0.33]$\,\, & & 74.666(1)\,\,\,\,$[0.38]$\,\, \\
\hline $\,0.24\,$ & $0.0037$ & 71.122(2)\,\,\,\,$[0.33]$\,\, & 71.141(1)\,\,\,\,$[0.37]$\,\, &
70.961(1)\,\,\,\,$[0.38]$\,\, \\
\hline $\,0.24\,$ & $0.0038$ & 69.411(2)\,\,\,\,$[0.32]$\,\, & & 69.254(1)\,\,\,\,$[0.37]$\,\, \\
\hline $\,0.30\,$ & $0.0040$ & 101.074(2)\,\,\,\,$[0.29]$\,\, & & 100.894(1)\,\,\,\,$[0.42]$\,\, \\
\hline $\,0.24\,$ & $0.0060$ & 46.025(2)\,\,\,\,$[0.29]$\,\, & & 45.929(1)\,\,\,\,$[0.35]$\,\, \\
\hline $\,0.24\,$ & $0.0080$ & 35.781(2)\,\,\,\,$[0.24]$\,\, & & 35.710(1)\,\,\,\,$[0.31]$\,\, \\
\hline
\end{tabular*}
\end{center}
\newpage
%
%
\begin{center}
\large TABLE 2 \small

\vspace{5mm}

\begin{tabular*}{\textwidth}[c]{@{\extracolsep{\fill}}|l||r|r|r||r|r|r||r|r|r||r|r|r|}
\hline $\,\,\,\,\,\,T^*$ & $P^*/T^*$ & $\rho_{\rm f}^*$ & $\rho_{\rm m}^*$ & $P^*/T^*$ & $\rho_{\rm BCC}^*$ & $\rho_{\rm
FCC}^*$ & $P^*/T^*$ & $\rho_{\rm FCC}^*$ & $\rho_{\rm BCC}^*$ & $P^*/T^*$ & $\rho_{\rm m}^*$ & $\rho_{\rm f}^*$\\
\hline\hline 0.0020 & 1.653 & 0.0779 & 0.0800 & -- & -- & -- & 20.360 & 0.1616 & 0.1618 & 650.20 & 0.6855 & 0.6863 \\
\hline 0.0030 & 2.146 & 0.0918 & 0.0938 & -- & -- & -- & 10.713 & 0.1494 & 0.1496 & & & \\
\hline 0.0033 & 2.324 & 0.0961 & 0.0978 & 2.520 & 0.1002 & 0.1003 & 8.397 & 0.1426 & 0.1427 & & & \\
\hline 0.0035 & 2.369 & 0.0980 & 0.0998 & 2.789 & 0.1046 & 0.1047 & 7.187 & 0.1382 & 0.1383 & & & \\
\hline 0.0037 & 2.537 & 0.1014 & 0.1031 & 3.510 & 0.1134 & 0.1135 & 5.784 & 0.1315 & 0.1316 & & & \\
\hline 0.0038 & 2.537 & 0.1020 & 0.1038 & -- & -- & -- & -- & -- & -- & & & \\
\hline 0.0040 & 2.730 & 0.1057 & 0.1074 & -- & -- & -- & -- & -- & -- & 185.90 & 0.5219 & 0.5230 \\
\hline 0.0060 & 3.893 & 0.1317 & 0.1332 & -- & -- & -- & -- & -- & -- & 78.670 & 0.4208 & 0.4220 \\
\hline 0.0080 & 7.285 & 0.1784 & 0.1792 & -- & -- & -- & -- & -- & -- & 31.172 & 0.3158 & 0.3166 \\
\hline
\end{tabular*}
\end{center}
\newpage
%
%
\begin{center}
\large TABLE 3 \normalsize

\vspace{5mm}

\begin{tabular*}{\textwidth}[c]{@{\extracolsep{\fill}}|c|c||l|l|l|}
\hline $1/n$ & $\rho^*$ & FCC\,\,\,($N=1372$) & HCP\,\,\,($N=1440$) & BCC\,\,\,($N=1458$) \\
\hline\hline $\,0.10\,$ & 2.5 & 50.350(1)\,\,\,$\left[1900\right]$ & & -- \\
\hline $\,0.12\,$ & 2.5 & 39.027(1)\,\,\,$\left[890\right]$ & & 39.397(2)\,\,\,$\left[340\right]$ \\
\hline $\,0.14\,$ & 3.0 & 47.700(1)\,\,\,$\left[850\right]$ & 47.708(1)\,\,\,$\left[870\right]$ &
47.943(2)\,\,\,$\left[490\right]$ \\
\hline $\,0.15\,$ & 3.5 & 59.180(1)\,\,\,$\left[1010\right]$ & & 59.423(2)\,\,\,$\left[600\right]$ \\
\hline $\,0.16\,$ & 4.0 & 69.608(1)\,\,\,$\left[1100\right]$ & 69.624(1)\,\,\,$\left[1110\right]$ &
69.833(2)\,\,\,$\left[700\right]$ \\
\hline $\,0.17\,$ & 3.5 & 50.607(1)\,\,\,$\left[570\right]$ & & 50.687(2)\,\,\,$\left[400\right]$ \\
\hline $\,0.20\,$ & 5.0 & 76.881(1)\,\,\,$\left[690\right]$ & 76.926(1)\,\,\,$\left[680\right]$ &
76.947(1)\,\,\,$\left[540\right]$ \\
\hline
\end{tabular*}
\end{center}
\newpage
%
%
\begin{center}
\large TABLE 4 \normalsize

\vspace{5mm}

\begin{tabular*}{\textwidth}[c]{@{\extracolsep{\fill}}|r||r|r|r||r|r|r|}
\hline $1/n$ & $P^*/T^*$ & $\rho_{\rm f}^*$ & $\rho_{\rm m}^*$ & $P^*/T^*$ & $\rho_{\rm BCC}^*$ & $\rho_{\rm FCC}^*$ \\
\hline\hline 0.10 & 29.747 & 1.289 & 1.327 & -- & -- & -- \\
\hline 0.12 & 41.810 & 1.501 & 1.534 & -- & -- & -- \\
\hline 0.14 & 60.507 & 1.787 & 1.817 & -- & -- & -- \\
\hline 0.15 & 71.151 & 1.942 & 1.968 & 76.945 & 2.020 & 2.024 \\
\hline 0.16 & 88.329 & 2.154 & 2.179 & 96.950 & 2.251 & 2.254 \\
\hline 0.17 & 105.614 & 2.362 & 2.387 & 124.717 & 2.534 & 2.537 \\
\hline 0.20 & 203.334 & 3.274 & 3.297 & 240.225 & 3.518 & 3.521 \\
\hline
\end{tabular*}
\end{center}

\newpage
%
%
\begin{figure}
\caption{\label{fig1}}
\includegraphics[width=16cm,angle=0]{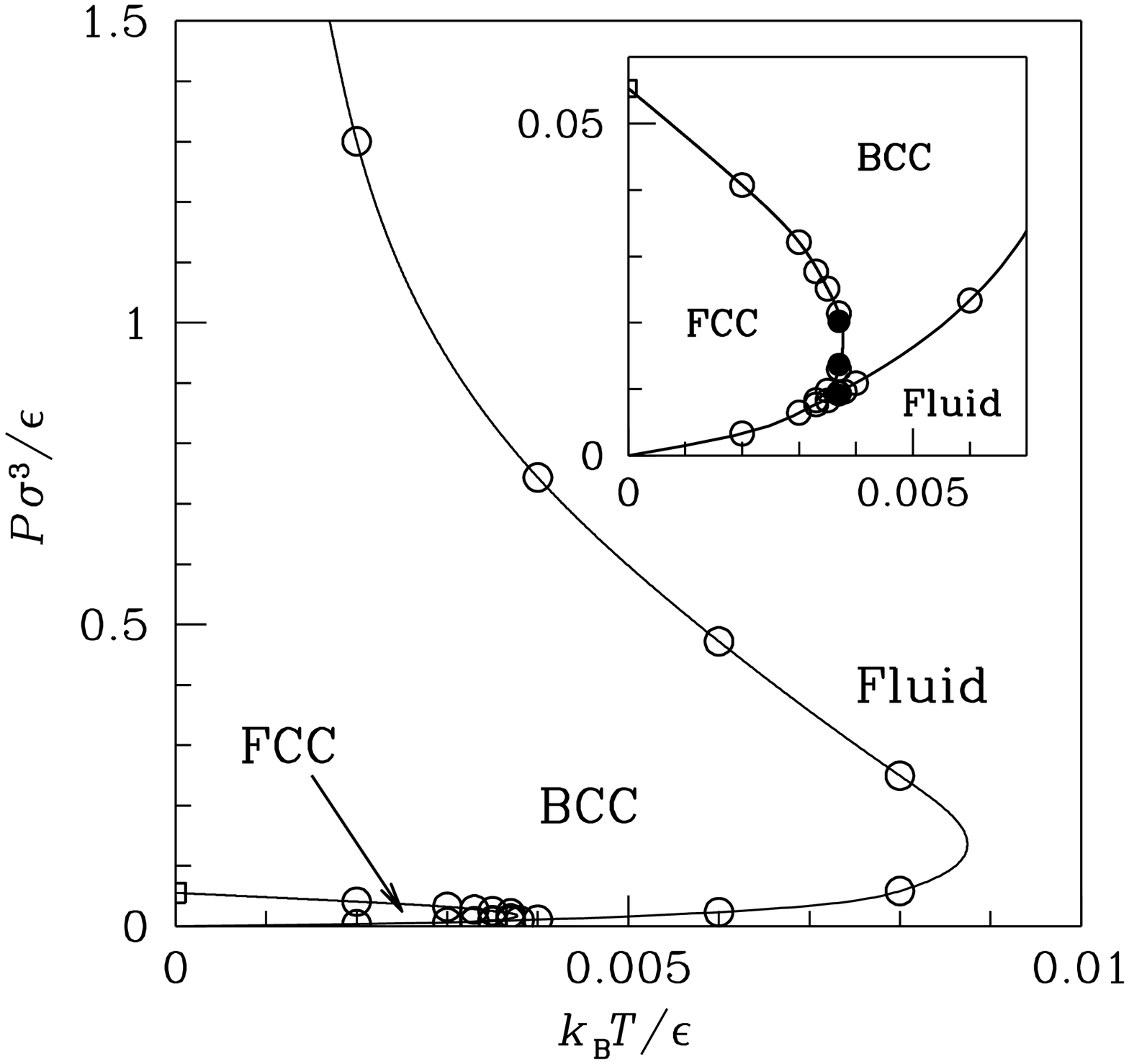}
\end{figure}
\newpage
%
%
\begin{figure}
\caption{\label{fig2}}
\includegraphics[width=16cm,angle=0]{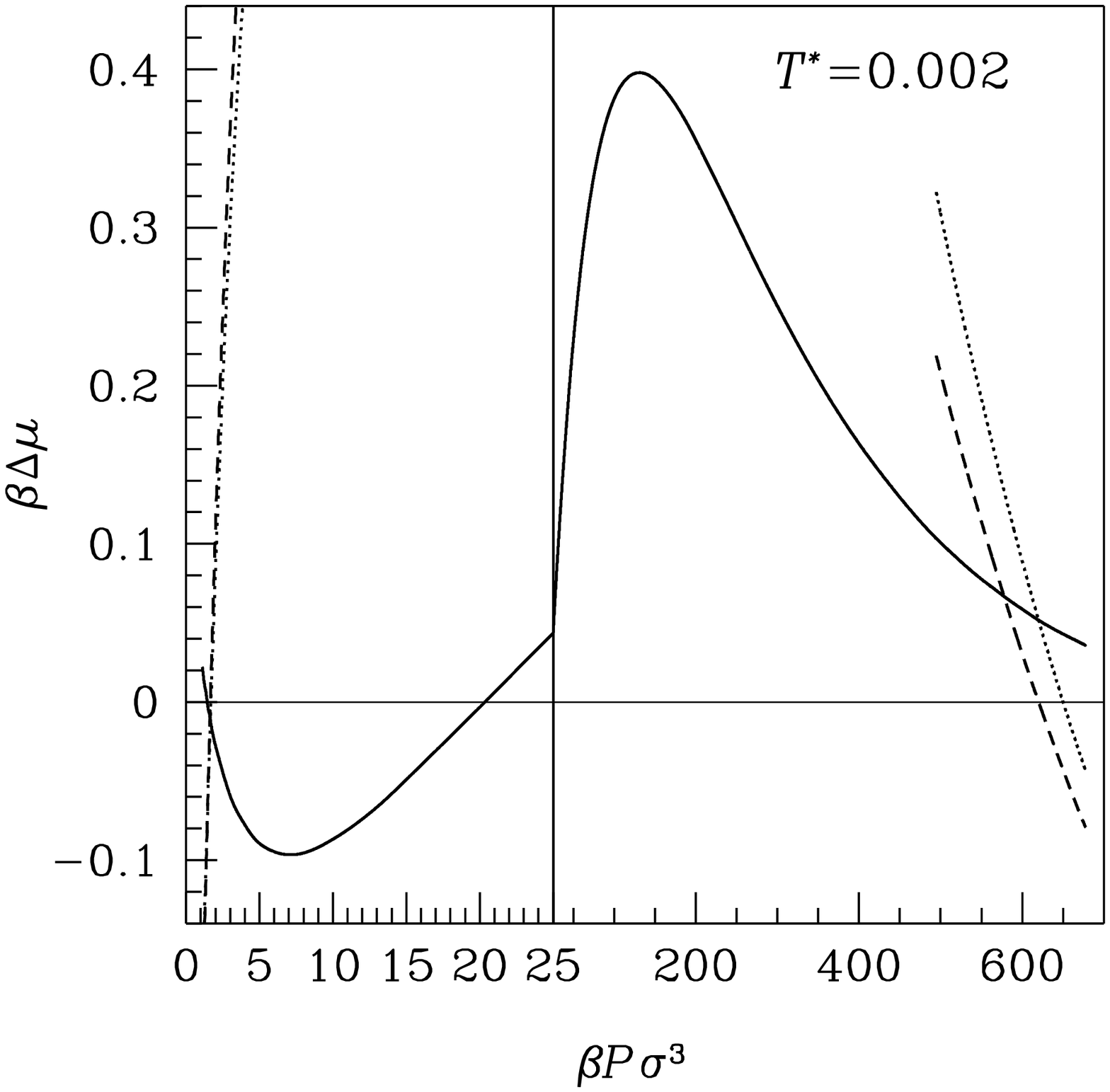}
\end{figure}
\newpage
%
%
\begin{figure}
\caption{\label{fig3}}
\includegraphics[width=16cm,angle=0]{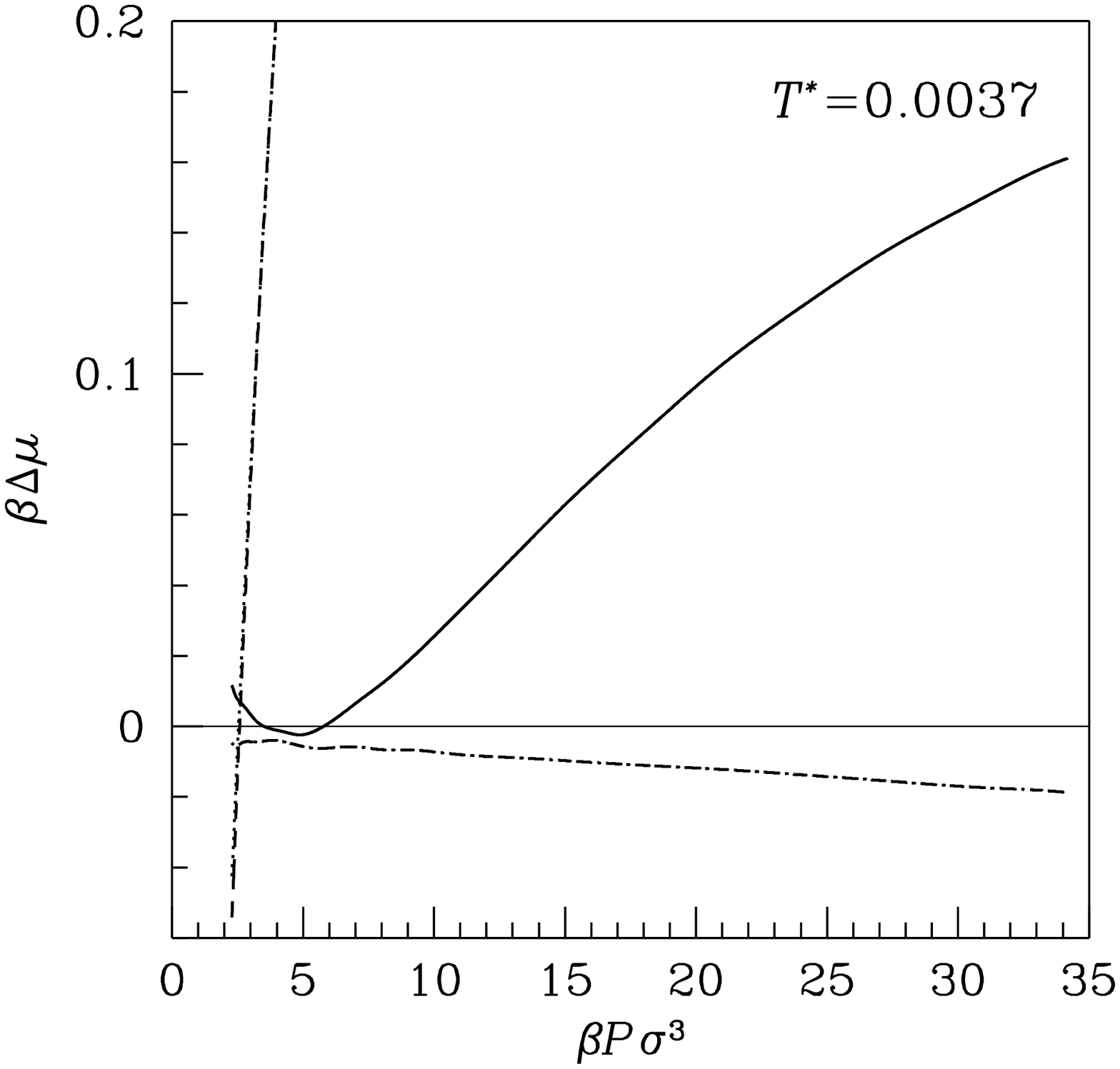}
\end{figure}
\newpage
%
%
\begin{figure}
\caption{\label{fig4}}
\includegraphics[width=16cm,angle=0]{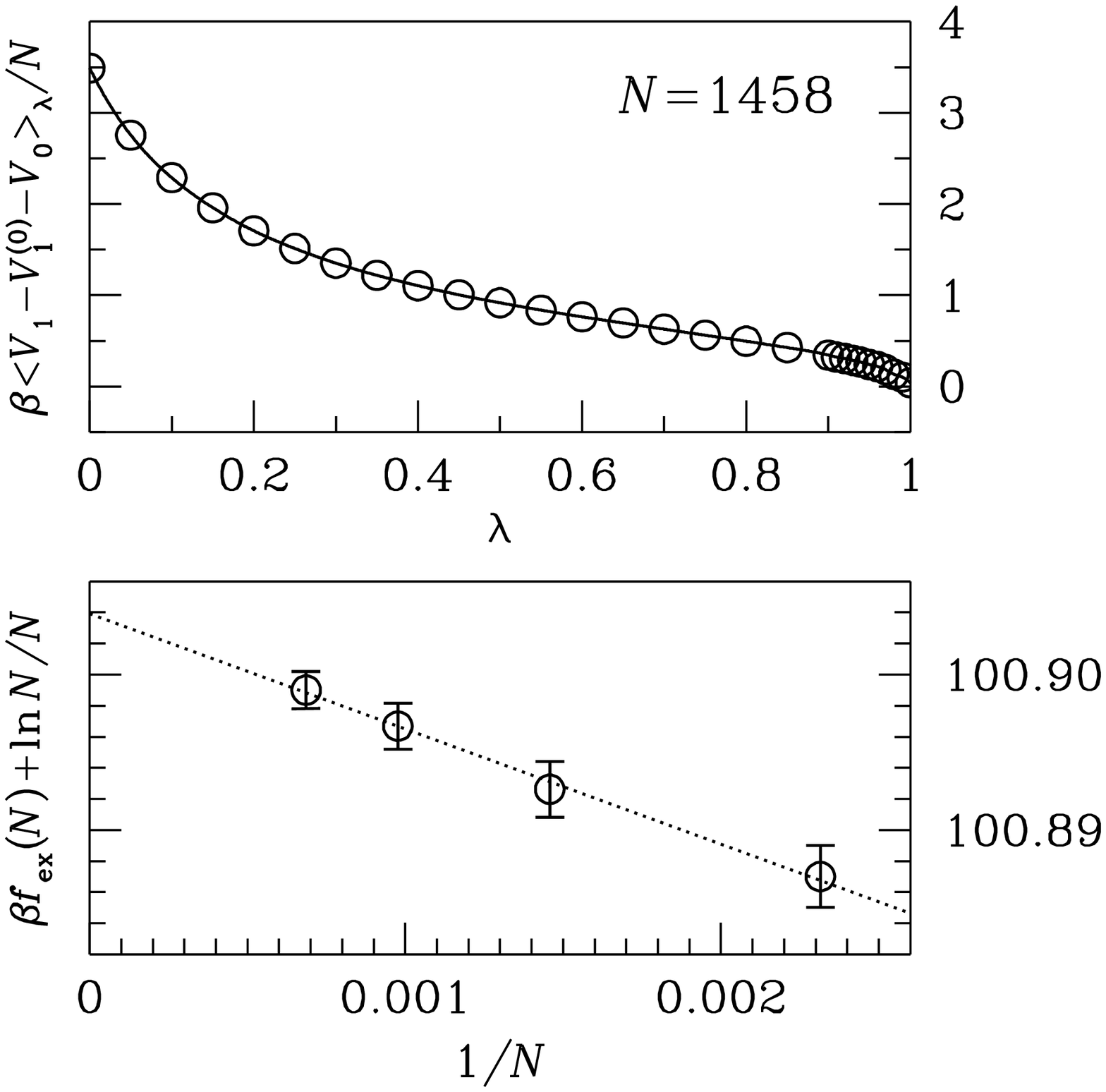}
\end{figure}
\newpage
%
%
\begin{figure}
\caption{\label{fig5}}
\includegraphics[width=16cm,angle=0]{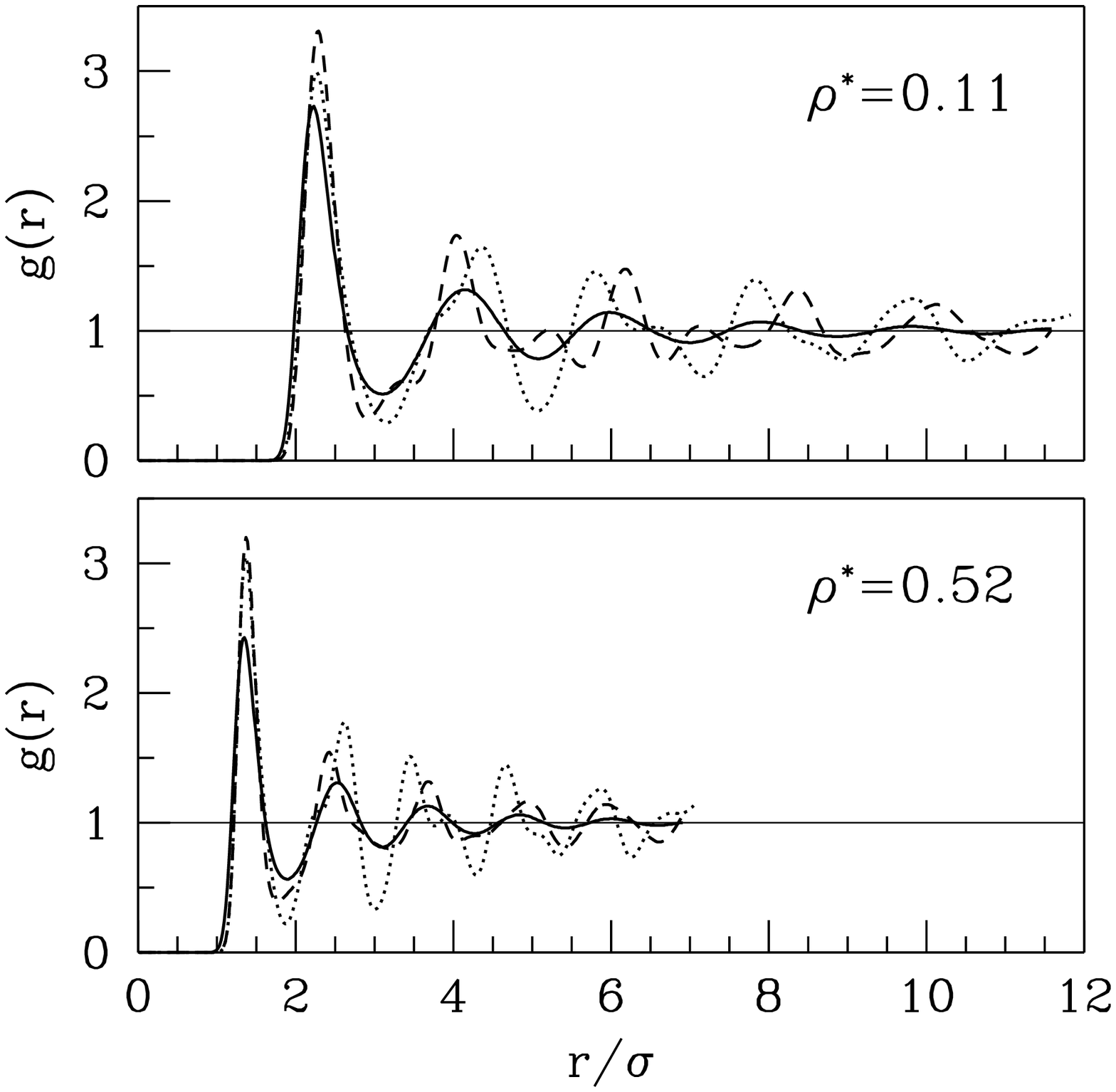}
\end{figure}
\newpage
%
%
\begin{figure}
\caption{\label{fig6}}
\includegraphics[width=16cm,angle=0]{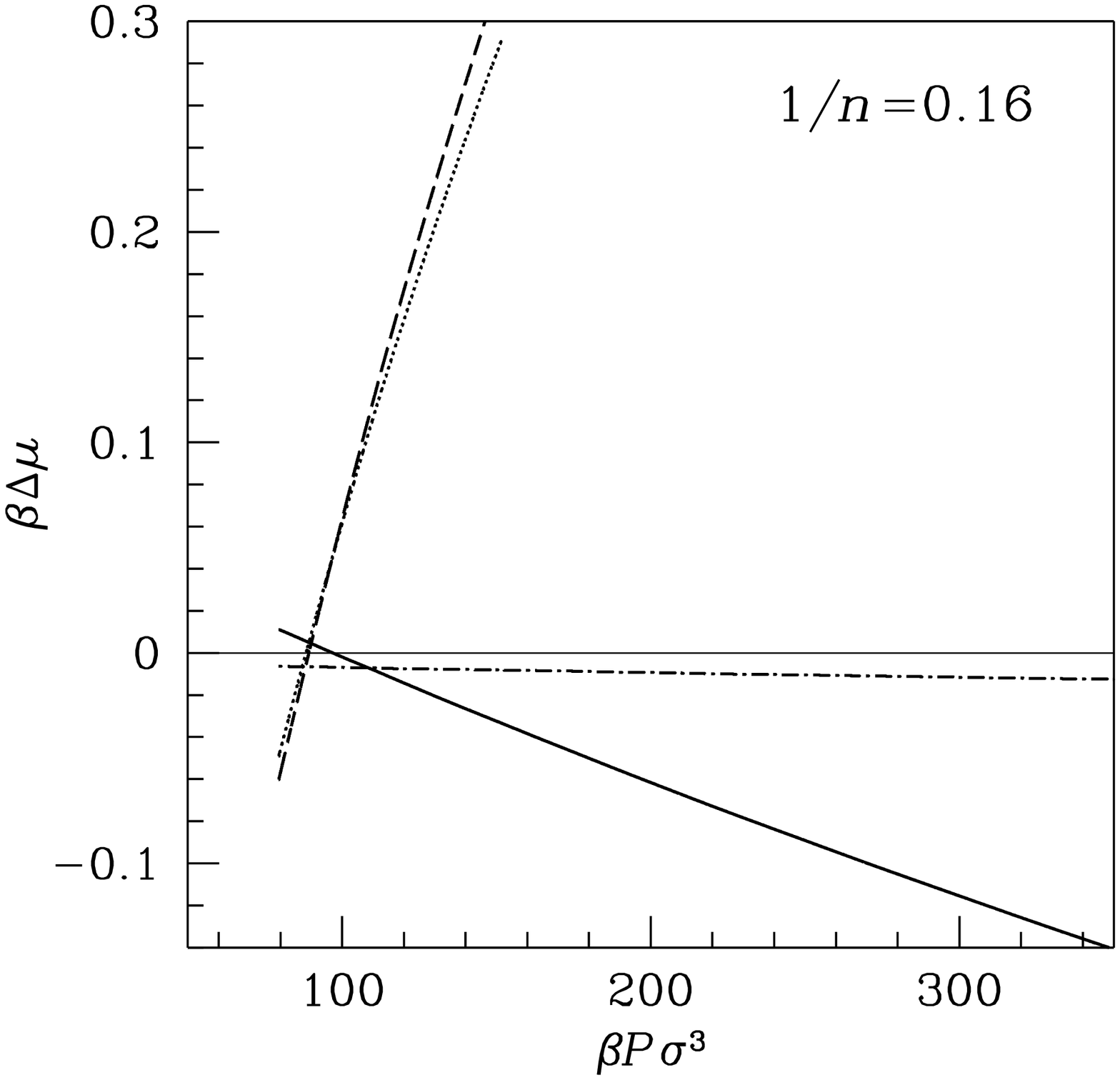}
\end{figure}
\newpage
%
%
\begin{figure}
\caption{\label{fig7}}
\includegraphics[width=16cm,angle=0]{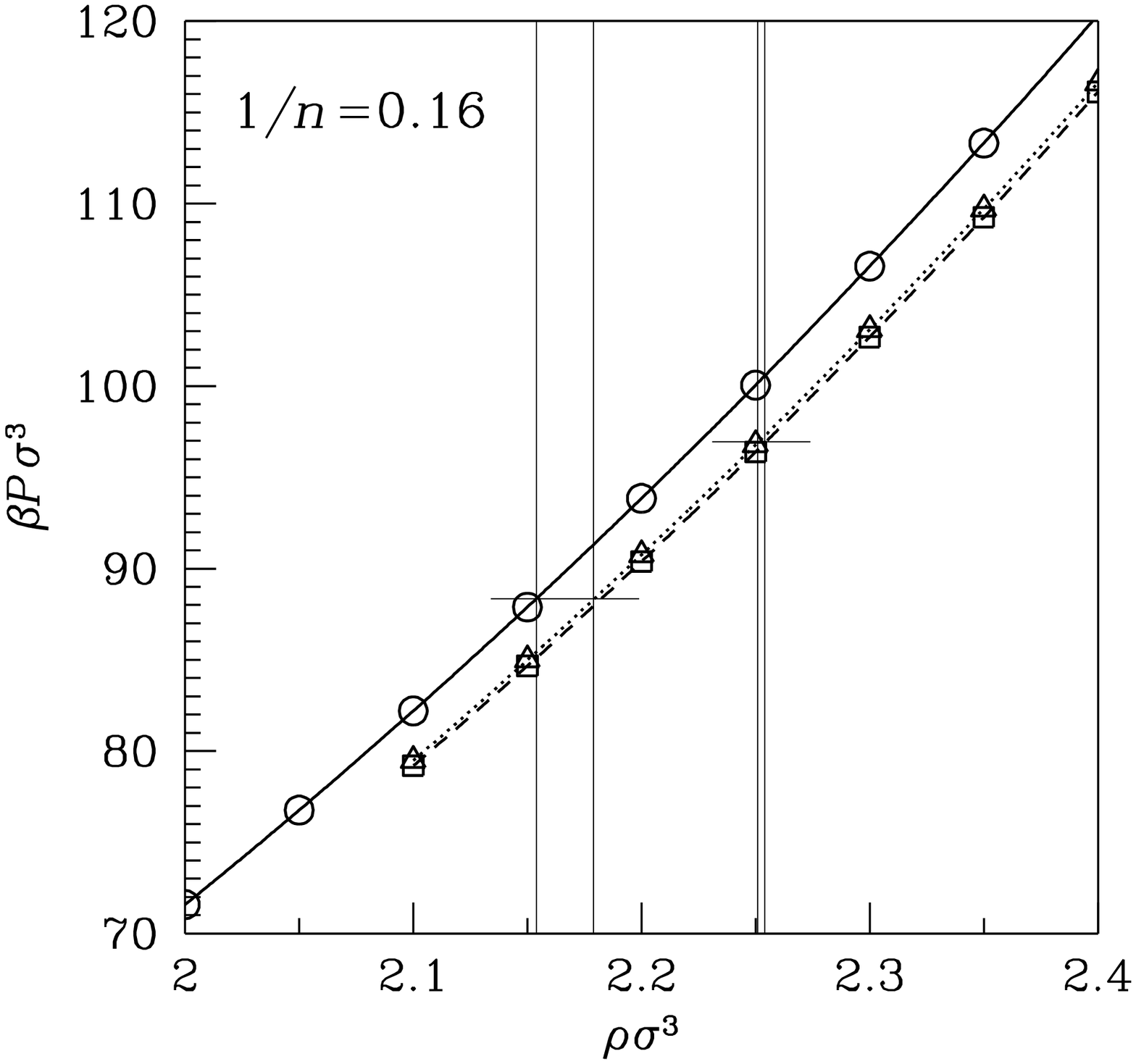}
\end{figure}
\newpage
%
%
\begin{figure}
\caption{\label{fig8}}
\includegraphics[width=16cm,angle=0]{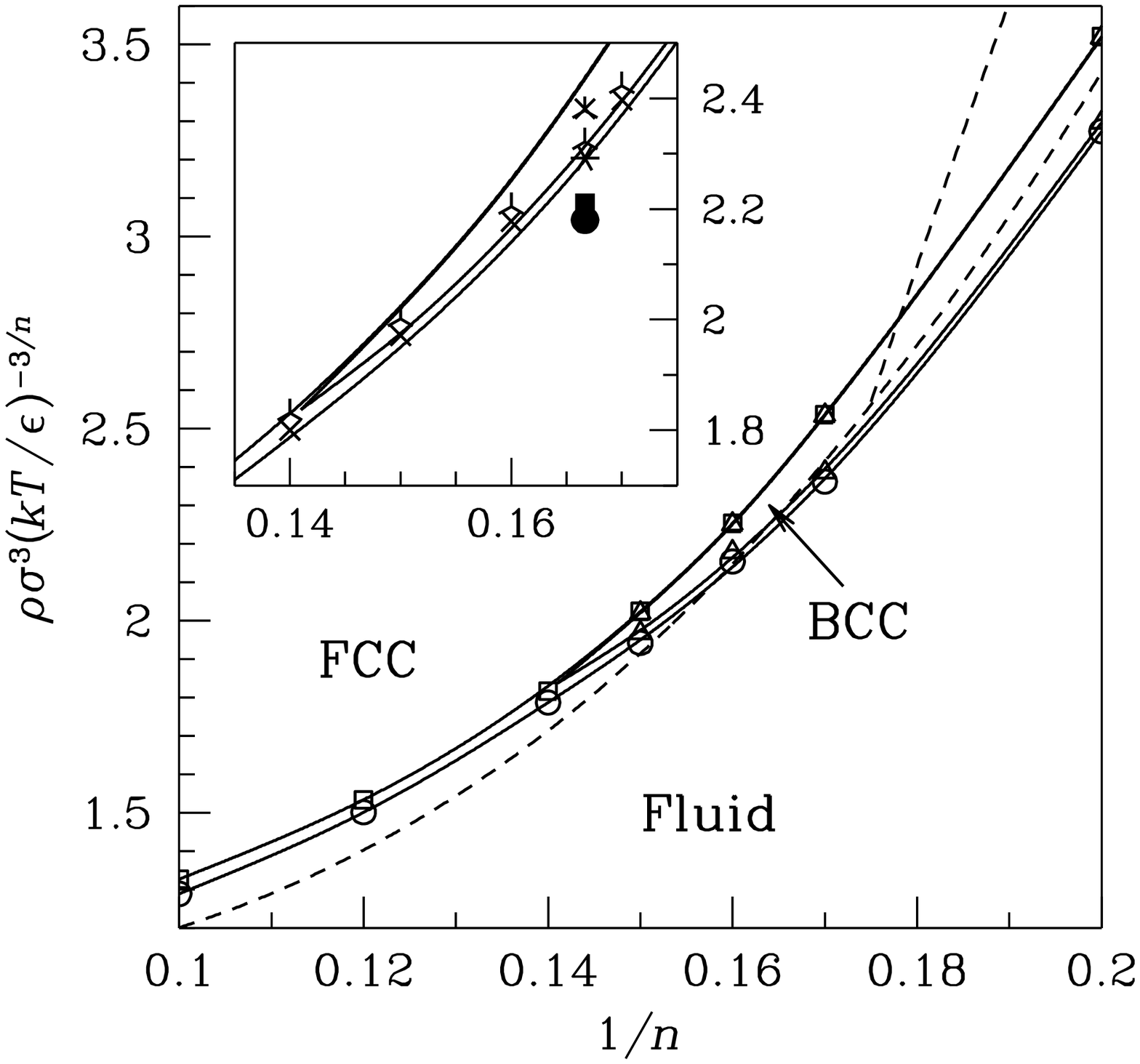}
\end{figure}

\begin{thebibliography}{99}
\bibitem{Pusey}  P.~N. Pusey, in {\em Liquids, Freezing, and the Glass Transition}, J.-P. Hansen {\it et al.} eds.
(North-Holland, Amsterdam, 1991).
\bibitem{Likos1}  C.~N. Likos, {\em Phys. Rep.} {\bf 348}, 267 (2001).
\bibitem{Louis}  A.~A. Louis, P.~G. Bolhuis, J.-P. Hansen, and E.~J. Meijer, {\em Phys. Rev. Lett.} {\bf 85}, 2522
(2000)
\bibitem{Bolhuis}  P.~G. Bolhuis, A.~A. Louis, J.-P. Hansen, and E.~J. Meijer, {\em J. Chem. Phys.} {\bf 114}, 4296
(2001).
\bibitem{Stillinger1}  F.~H. Stillinger, {\em J. Chem. Phys.} {\bf 65}, 3968 (1976).
\bibitem{Hoover1}  W.~G. Hoover, D.~A. Young, and R. Grover, {\em J. Chem. Phys.} {\bf 56}, 2207 (1972).
\bibitem{Vocadlo}  L. Vo\u{c}adlo and D. Alf\`e, {\em Phys. Rev. B} {\em 65}, 214105 (2002).
\bibitem{Frenkel1}  D. Frenkel and A.~J.~C. Ladd, {\em J. Chem. Phys.} {\bf 81}, 3188 (1984).
\bibitem{Polson}  J.~M. Polson, E. Trizac, S. Pronk, and D. Frenkel, {\em J. Chem. Phys.} {\bf 112}, 5339 (2000).
\bibitem{Frenkel2}  D. Frenkel and B. Smit, {\em Understanding Molecular Simulation} (Academic Press, 2001).
\bibitem{Meijer}  E.~J. Meijer and D. Frenkel, {\em J. Chem. Phys.} {\bf 94}, 2269 (1991).
\bibitem{Hamaguchi}  S. Hamaguchi, R.~T. Farouki, and D.~H.~E. Dubin, {\em Phys. Rev. E} {\bf 56}, 4671 (1997).
\bibitem{Robbins}  M.~O. Robbins, K. Kremer, and G.~S. Grest, {\em J. Chem. Phys.} {\bf 88}, 3286 (1988).
\bibitem{Sirota} E.~B. Sirota, H.~D. Ou-Yang, S.~K. Sinha, and P.~M. Chaikin, {\em Phys. Rev. Lett.} {\bf 62}, 1524
(1989).
\bibitem{Witten}  T.~A. Witten and P.~A. Pincus, {\em Macromolecules} {\bf 19}, 2509 (1986).
\bibitem{Likos2}  C.~N. Likos, H. L\"{o}wen, M. Watzlawek, B. Abbas, O. Jucknischke, J. Allgaier, and D. Richter, {\em
Phys. Rev. Lett.} {\bf 80}, 4450 (1998).
\bibitem{Watzlawek}  M. Watzlawek, C.~N. Likos, and H. L\"owen, {\em Phys. Rev. Lett.} {\bf 82}, 5289 (1999).
\bibitem{Prestipino}  S. Prestipino, F. Saija, and P.~V. Giaquinta, {\em Phys. Rev. E} {\bf 71},  050102(R) (2005).
\bibitem{Likos3}  C.~N. Likos, M. Watzlawek, and H. L\"owen, {\em Phys. Rev. E} {\bf 58}, 3135 (1998).
\bibitem{Schmidt}  M. Schmidt, {\em J. Phys.: Condens. Matter} {\bf 11}, 10163 (1999).
\bibitem{Stillinger2}  F.~H. Stillinger, {\em Phys. Rev. B} {\bf 20}, 299 (1979).
\bibitem{Stillinger3}  F.~H. Stillinger and D.~K. Stillinger, {\em Physica A} {\bf 244}, 358 (1997).
\bibitem{Lang}  A. Lang, C.~N. Likos, M. Watzlawek, and H. L\"owen, {\em J. Phys.: Condens. Matter} {\bf 12}, 5087
(2000).
\bibitem{Likos4}  C.~N. Likos, A. Lang, M. Watzlawek, and H. L\"{o}wen, {\em Phys. Rev. E} {\bf 63}, 031206 (2001).
\bibitem{Hoover2}  W.~G. Hoover, M. Ross, K.~W. Johnson, D. Henderson, J.~A. Barker, and B.~C. Brown, {\em J. Chem.
Phys.} {\bf 52 }, 4931 (1970); W.~G. Hoover, S.~G. Gray, and K.~W. Johnson, {\em J. Chem. Phys.} {\bf 55}, 1128
(1971).
\bibitem{Laird} B.~B. Laird and A.~D.~J. Haymet, {\em Mol. Phys.} {\bf 75}, 71 (1992).
\bibitem{Agrawal} R. Agrawal and D.~A. Kofke, {\em Phys. Rev. Lett.} {\bf 74}, 122 (1995); {\em Mol. Phys.} {\bf 85}, 23
(1995).
\bibitem{Brush}  S.~G. Brush, H.~L. Sahlin, and E. Teller, {\em J. Chem. Phys.} {\bf 45}, 2102 (1966).
\bibitem{Hansen}  J.-P. Hansen, {\em Phys. Rev. A} {\bf 8}, 3096 (1973).
\bibitem{Widom}  B. Widom, {\em J. Chem. Phys.} {\bf 39}, 2808 (1963).
\bibitem{Stillinger4}  M.~R. Feeney, P.~G. Debenedetti, and F.~H. Stillinger {\em J. Chem. Phys.} {\bf 119}, 4582
(2003).
\bibitem{Giaquinta}  P.~V. Giaquinta and F. Saija, {\em Chem. Phys. Chem.} (2005), Issue 9.
\bibitem{Wang}  D.~C. Wang and A.~P. Gast, {\em J. Chem. Phys.} {\bf 110}, 2522 (1999).
\bibitem{Dubin}  D.~H.~E. Dubin and H. Dewitt, {\em Phys. Rev. B} {\bf 49}, 3043 (1994).
\bibitem{Saija}  F. Saija and S. Prestipino, {\em Phys. Rev. B} {\bf 72}, 024113 (2005).
\end{thebibliography}
\end{document}